# Quantum Entanglement and Cryptography for Automation and Control of Dynamic Systems

Farbod KHOSHNOUD[1,2], Ibrahim I. ESAT[3], Shayan JAVAHERIAN[4], Behnam BAHR[4]

[1]*Department of Electromechanical Engineering Technology, College of Engineering, California State Polytechnic University, Pomona, CA 91768, USA*

[2]*Center for Autonomous Systems and Technologies, Department of Aerospace Engineering, California Institute of Technology, Pasadena, CA 91125, USA*

[3]*Department of Mechanical and Aerospace Engineering, Brunel University London, Uxbridge UB8 3PH, United Kingdom*

[4]*Department of Mechanical Engineering, College of Engineering, California State Polytechnic University, Pomona, CA 91768, USA*

**Abstract:** This paper addresses the application of quantum entanglement and cryptography for automation and control of dynamic systems. A dynamic system is a system where the rates of changes of its state variables are not negligible. Quantum entanglement is realized by the Spontaneous Parametric Down-conversion process. Two entangled autonomous systems exhibit correlated behavior without any classical communication in between them due to the quantum entanglement phenomenon. Specifically, the behavior of a system, Bob, at a distance, is correlated with a corresponding system, Alice. In an automation scenario, the "Bob Robot" is entangled with the "Alice Robot" in performing autonomous tasks without any classical connection between them. Quantum cryptography is a capability that allows guaranteed security. Such capabilities can be implemented in control of autonomous mechanical systems where, for instance, an "Alice Autonomous System" can control a "Bob Autonomous System" for applications of automation and robotics. The applications of quantum technologies to mechanical systems, at a scale larger than the atomistic scale, for control and automation, is a novel contribution of this paper. Notably, the feedback control transfer function of an integrated classical dynamic system and a quantum state is proposed.

# 1 Introduction

In this Quantum era, quantum technologies are making the biggest and the most promising impacts in developing new technologies towards improving the quality of life. The applications of quantum technologies in our future everyday lives are inevitable due to their many unmatched advantages. By integrating the unmatched possibilities of quantum supremacy with engineering applications, such interdisciplinary quantum and engineering systems can push the engineering boundaries beyond any classical technique. However, currently, the most focus on the application of quantum technologies in engineering systems is on quantum computing and related areas (For example, see [1], [4]). Until recently, a Quantum Robot is referred to as a mobile autonomous platform that is equipped with a quantum computer, as its processing system (See for example, [3]). Also, the advantages of quantum technologies have not been realized and introduced at the macro scale. Very few references can be found on the application of quantum capabilities to actual mechanical systems. The applications of quantum technologies in advancing the performance of mechanical systems (at the macro-scale) are found in the literature merely on developing novel sensors and actuators ([5]-[15]), and quantum information processing systems, with optics-based distributed networks (See for example, [16]), optical switching, and a self-correcting quantum memory, for developing quantum devices that can control themselves [16].

The applications of quantum optics in controls [17]-[20], feedback systems [21]-[22], and programmable logic devices in quantum optics [23] have shown significant advances. An integration of experimental quantum interference with soft computing as an experimental quantum-enhanced stochastic simulation device can execute a simulation using less memory than possible by classical means [24].

Although there is rich literature in areas related to "quantum engineering" (e.g., [16]-[24]), the actual integration of such technologies with macro-scale mechanical systems, for instance in robotics, incorporation of autonomy, and autonomous systems (e.g., unmanned systems), has not been addressed, despite its importance and its capacity for potential technological breakthrough.

Optics technologies are used in quantum mechanics, where single photons (or alternatively electrons, or sound energy levels [25]) are considered in the study of a quantum system. Similar, already well-developed technologies in optics and photonics areas are utilized in quantum mechanics. For instance, classical free-space optical communication tools in the autonomy of unmanned systems [26] (although not in quantum context), can be implemented towards the development of quantum engineering systems.

The applications of the principles of quantum mechanics integrated with the laws of analytical mechanics in analyzing the motion of elementary particles (for example, see [27]), such as molecular dynamics topics, are readily available. However, the problem of multibody dynamical systems (at the macro-scale) integrated with quantum mechanics (for instance, where the quantum control of the system is of interest) has not been addressed yet. The research by the authors of this paper, and their earlier work (see for example, [48]-[50]) is the first effort towards the theoretical and experimental research and establishing the interdisciplinary field of Quantum Multibody Dynamics ([28], [48]-[50]).

Recent advances in experimental and theoretical quantum systems allows: Secure quantum communication code, with no classical communication [29], quantum correlations over more than 10 km [30], Entanglement in noisy quantum channels [31], Entanglement-based quantum communication over 144 km [32], free-space entangled photon distributions in long distances [33], and Secured quantum key distribution with entangled photons [34]. A quantum-inspired approach has been proposed to solve problems of two robotic agents finding each other or pushing an object ([35], [36]), without any knowledge of each other. Furthermore, a research work on Psi Intelligent Control [37], inspired by precognition, has led to an approach in a quantum entanglement-based autonomy concept for autonomous vehicles [38].

Recent technological advances in quantum mechanics make such experiments feasible with considerable cost and size reductions [39]-[44]. On the other hand, reconfigurable Quantum Key Distribution networking [45] techniques allow free-space quantum communication over significant distances and overcomes signal degradation issues (e.g., due to weather events), where, unconditionally secure bit commitment by transmitting measurement outcomes is possible to attain perfect security ([46], [47]). Collectively, such technology resources and potential capacities allow us to apply quantum capabilities in engineering applications effectively, particularly in robotics and autonomous domains.

Developing quantum capabilities in the mechanical system domain pushes the engineering boundaries beyond any existing classical technique, which gives advantageous and unmatched capabilities, such as the possibility of entangling the robotic agents in a distributed robotic system, quantum superposition, and guaranteed security. The present paper introduces the integration of quantum technologies with engineering systems (in a physical system domain), where the quantum advantages are applied to robots and autonomous systems in cooperative multi-agent robotic network scenarios. It demonstrates how experimental quantum entanglement and cryptography can be integrated with engineering systems (for example, how quantum can be used in control of a mechanical system). To the authors' knowledge, this paper, and their recent work ([48]-[50]), are the first attempts in establishing the integration of experimental quantum capabilities with multibody dynamic systems (such as multi-agent robotic problems). A goal for such integrated quantum engineering approach is to demonstrate the interface of Quantum Technologies with Engineering Systems for research and educational purposes. The authors introduced the concept of experimental quantum cryptography and entanglement in robotics for the first time in the references [48]-[50]. The concepts of Quantum Multibody Dynamics, Robotics, Controls, and Autonomy are integrated with, and take advantage of, quantum technologies here. The paper is organized as follows. Section 2 presents an automation scenario, and Section 3 addresses the classical dynamics and controls aspects of such problems. Sections 4-6 give an introduction to quantum mechanics (to allow the readers to use the paper as a self-contained article). In Sections 7, the control problem for autonomy in the context of quantum states as the integrated quantum and classical feedback control transfer function for the autonomy of mechanical systems, at nonatomistic scales, is proposed, for the first time. Sections 8 and 9 discuss the quantum cryptography and quantum entanglement, respectively, and how they are implemented for automation and robotics applications.

## 2 Automation

The aim of the present paper to introduce a technique to enhance autonomy through quantum technologies. An integrated automation and robotics system is shown in Fig. 1. This system consists of autonomous mobile platforms, robotic manipulators, and an automation facility equipped with actuators, motion detection sensors, and microcontrollers. The autonomy of such systems enhanced by quantum entanglement and quantum cryptography, is discussed in the following sections.

## 3 Dynamics of the Autonomous Platforms

The equations of motion of an autonomous system containing robotic platforms and manipulators are presented now. A schematic representation of the system is shown in Fig. 2. The moment of inertia matrix of the vehicle in principal axes is given by,

$$I_{inertia} = \begin{bmatrix} I_{xx} & 0 & 0 \\ 0 & I_{yy} & 0 \\ 0 & 0 & I_{zz} \end{bmatrix}$$

The Euler angles for the sequence of rotations from the inertial frame, $I$, to the body frame, $B$, can be given by yaw, then pitch, and then roll, with yaw rotation $\theta_z$ about $z_I$, in the inertial frame, or in mathematical form $r_1 = \mathbf{H}_I^1 r_I$, pitch rotation $\theta_y$ about $y_1$, in the intermediate frame, or $r_2 = \mathbf{H}_1^2 r_1$, and roll rotation $\theta_x$ about $x_2$, in the body frame, as $r_B = \mathbf{H}_2^B(\theta_x) r_2$.

The three-angle rotation matrix is the product of 3 single-angle rotation matrices, as,

$$\mathbf{H}_I^B = \mathbf{H}_2^B(\theta_x)\mathbf{H}_1^2(\theta_y)\mathbf{H}_I^1(\theta_z)$$

where $\mathbf{H}_I^B(\theta_x, \theta_y, \theta_z)$ is the rotation matrix of the inertial frame with respect to the body frame, and $[\mathbf{H}_I^B]^{-1} = [\mathbf{H}_I^B]^T = \mathbf{H}_B^I$; and $\mathbf{H}_B^I \mathbf{H}_I^B = \mathbf{H}_I^B \mathbf{H}_B^I = \mathbf{I}$.

The relationship between the Euler-angle rates and body-axis rates is expressed as,

$$\begin{bmatrix} p \\ q \\ r \end{bmatrix} = \begin{bmatrix} 1 & 0 & -\sin\theta_y \\ 0 & \cos\theta_x & \sin\theta_x \cos\theta_y \\ 0 & -\sin\theta_x & \cos\theta_x \cos\theta_y \end{bmatrix} \begin{bmatrix} \dot{\theta}_x \\ \dot{\theta}_y \\ \dot{\theta}_z \end{bmatrix} = \mathbf{L}_I^B \dot{\Theta}$$

Or

$$\begin{bmatrix} \dot{\theta}_x \\ \dot{\theta}_y \\ \dot{\theta}_z \end{bmatrix} = \mathbf{L}_B^I \boldsymbol{\omega}_B$$

where $\dot{\theta}_z$ is measured in the inertial frame, $\dot{\theta}_y$ is measured in the intermediate frame, and $\dot{\theta}_x$ is measured in the body frame. Here, $\dot{\Theta}$ denotes the Euler angle rate vector, and $\boldsymbol{\omega}_B$ is the angular velocity vector in the body frame.

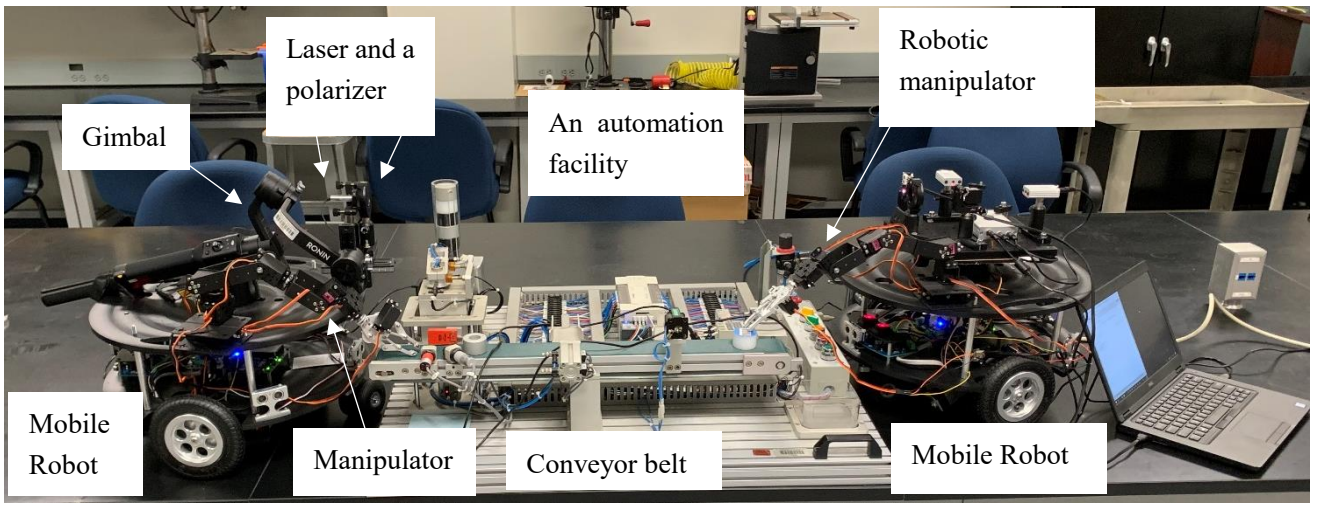

**Fig. 1 An integrated automation and robotic system.**

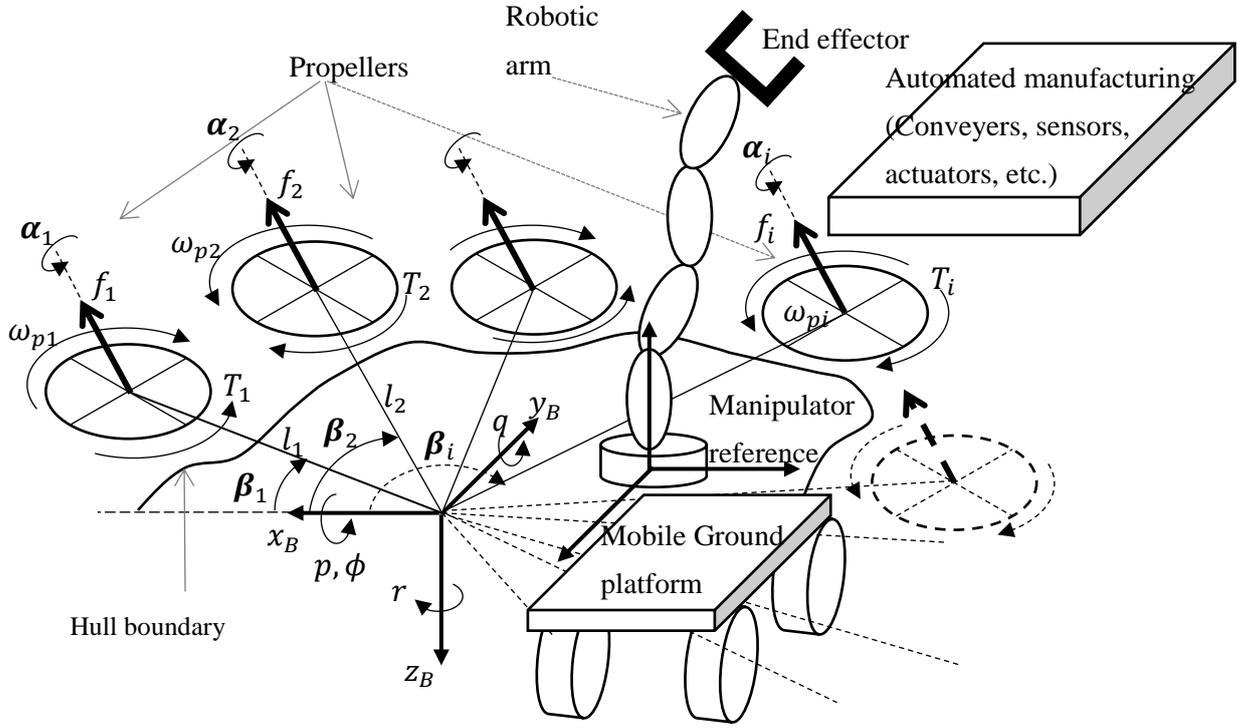

**Fig. 2 A generalized automation system and robotic platform.**

The rate of change of the translational position is obtained by

$$\dot{\mathbf{r}}_I(t) = \mathbf{H}_B^I(t)\boldsymbol{v}_B(t)$$

where $\mathbf{r}_I = \begin{bmatrix} x \\ y \\ z \end{bmatrix}_I$ is the translational position, and $\boldsymbol{v}_B = \begin{bmatrix} u \\ v \\ w \end{bmatrix}$ is the translational velocity in the body frame.

The rate of change of angular position is
$$\dot{\boldsymbol{\Theta}}(t) = \mathbf{L}_B^I(t)\boldsymbol{\omega}_B(t)$$

The equaion of motion can be written in terms of the rate of change of translational velocity as,

$$\dot{\boldsymbol{v}}_B(t) = \frac{1}{m(t)}\mathbf{F}_B(t) + \mathbf{H}_I^B(t)\mathbf{g}_I - \widetilde{\boldsymbol{\omega}}_B(t)\boldsymbol{v}_B(t) \quad (1)$$

and the rate of change of angular velocity is,

$$\dot{\boldsymbol{\omega}}_B(t) = \mathbf{I}_B^{-1}(t)[\mathbf{M}_B(t) - \widetilde{\boldsymbol{\omega}}_B(t)\mathbf{I}_B(t)\boldsymbol{\omega}_B(t)] \quad (2)$$

where $\widetilde{\boldsymbol{\omega}}$ is the cross product equivalent matrix of $\boldsymbol{\omega}$ given by,

$$\widetilde{\boldsymbol{\omega}} = \begin{bmatrix} 0 & -\omega_z & \omega_y \\ \omega_z & 0 & -\omega_x \\ -\omega_y & \omega_x & 0 \end{bmatrix}$$

and $\dot{\mathbf{H}}_I^B = -\widetilde{\boldsymbol{\omega}}_B \mathbf{H}_I^B$; $\dot{\mathbf{H}}_B^I = \widetilde{\boldsymbol{\omega}}_I \mathbf{H}_B^I$.

The external forces include the drag and propulsion forces in the body frame as,

$$\mathbf{F}_B = \mathbf{F}_{B,drag} + \mathbf{F}_{B,propul} = \begin{bmatrix} X_{drag} + X_{propul} \\ Y_{drag} + Y_{propul} \\ Z_{drag} + Z_{propul} \end{bmatrix}_B \quad (3)$$

and the applied moments in body frame are

$$\mathbf{M}_B = \mathbf{M}_{B,drag} + \mathbf{M}_{B,propul} \quad (4)$$
$$= \begin{bmatrix} L_{drag} + L_{propul} \\ M_{drag} + M_{propul} \\ N_{drag} + N_{propul} \end{bmatrix}_B$$

The propulsion forces are generalized here, which can represent the driving force of a ground vehicle and the thrust of the propellers of an aerial vehicle [51], [52]. In the case of a mobile ground platform, as in Fig. 2, the vertical translations are due to the road roughness.

The motion of the end effector in Fig. 2 with respect to the reference frame, $R$ (located at the base of the manipulator), can be obtained by,

$$^R T_{E_{eff}} = {}^R T_1 \, {}^1 T_2 \, {}^2 T_3 \, \cdots \, {}^{fn-1} T_{fn} \, {}^{fn} T_{E_{eff}}$$

where $^{rf}T_E$ is the transformation matrix of frame $E$ relative to frame $fn$, and $fn$ denotes the frame of link $n$.

If the position and orientation of the end effector are given by the vectors $\boldsymbol{P}_E$ and $\boldsymbol{O}_E$ relative to the reference frame $R$, attached to the mobile vehicle, we have,

$$\begin{bmatrix} \boldsymbol{P}_E \\ \boldsymbol{O}_E \end{bmatrix}_I = \begin{bmatrix} \boldsymbol{P}_E \\ \boldsymbol{O}_E \end{bmatrix}_R + \begin{bmatrix} \mathbf{r}_I \\ \boldsymbol{\Theta}_I \end{bmatrix}$$

The control of the robotic platforms in a multi-agent networked scenario is discussed now. In a multi-agent robotic system scenario, various control tasks may be desirable, such as formations, swarm and cooperative physical tasks. In performing robotic tasks, a common problem is involved with the kinematic synchronization of the robotic platforms in a network of autonomous systems, in a dynamic environment. For instance, in a formation problem for a group of robotic platforms (or unmanned autonomous systems), maintaining the reference relative positions between the agents is required. Such formation may be defined by,

$$\lim_{t \to \infty} \left\| \left( \mathbf{r}_{I,i}(t) - \mathbf{r}_L(t) \right) - \mathbf{d}_i(t) \right\| = 0$$

where $\mathbf{r}_{I,i}$ is the position of the robotic agent $i$, $\mathbf{r}_L$ is the position of a reference agent, and $\mathbf{d}_i$ is the desired position vector of robot $i$ with respect to the reference or another agent. A controller $g(k_{pid})$ can now be implemented to realize the required task, by controlling the actuation forces of the individual robots that drive each agent $i$, with a strategy given by,

$$\sum_{j=1}^{N+1} a_{ij} \left\{ g(k_{pid}) \left[ \left( \mathbf{r}_{I,i}(t) - \mathbf{r}_{I,j}(t) \right) - \mathbf{d}_i(t) \right] \right\}$$

where $a_{ij}$ is equal to 1 if an agent $i$ is connected to $j$, and it is equal to zero otherwise.

## 4 The Quantum State Vector

The spin state of a particle is denoted by $|+\mathbf{z}\rangle$ with the value of intrinsic angular momentum in the $z$ direction, $S_z$, equal to $\hbar/2$, where $\hbar = h/2\pi = 1.055 \times 10^{-27}$ erg · s $= 6.582 \times 10^{-16}$ eV · s, and $h$ is the Planck's constant. In classical physics, a coordinate system with the basis vectors $\mathbf{i}$, $\mathbf{j}$, and $\mathbf{k}$ is applied in representing vectors. The quantum representation of vectors is realized based on the assumption that if a (spin-$\frac{1}{2}$) particle is sent to a magnetic field in the z direction we obtain only the values $\hbar/2$ and $-\hbar/2$, corresponding to the particle ending up in the state $|+\mathbf{z}\rangle$, and ending down in the state $|-\mathbf{z}\rangle$, respectively. These two states can be considered as the vectors in the quantum mechanical vector space. In this vector system, $\langle -\mathbf{z}|+\mathbf{z}\rangle = 0$ implies that a particle in the state $|+\mathbf{z}\rangle$ has an amplitude of zero to be in the state $|-\mathbf{z}\rangle$. This is analogous to the orthogonal property in classical physics (e.g., for an electric field, or a force field) where, for instance, $\mathbf{i} \cdot \mathbf{j} = 0$. Similarly, the amplitude of a particle in the state $|+\mathbf{z}\rangle$ to be in the state $|+\mathbf{z}\rangle$ is equal to one, or $\langle +\mathbf{z}|+\mathbf{z}\rangle = 1$ (the analogy is $\mathbf{i} \cdot \mathbf{i} = 1$ in classical Physics. If a particle initially in the state $|+\mathbf{x}\rangle$ passes through a magnetic field in the z direction, it can attain the amplitudes $c_+$ and $c_-$ in the states $|+\mathbf{z}\rangle$ and $|-\mathbf{z}\rangle$, respectively. Therefore, the state $|+\mathbf{x}\rangle$ can be expressed as a linear combination of the states $|+\mathbf{z}\rangle$ and $|-\mathbf{z}\rangle$ by,

$$|+\mathbf{x}\rangle = c_+|+\mathbf{z}\rangle + c_-|-\mathbf{z}\rangle \tag{5}$$

Assume an arbitrary spin state $|\psi\rangle$, created by sending a beam of particles with the intrinsic spin-$\frac{1}{2}$ through an inhomogeneous magnetic field oriented in some arbitrary direction. If the measurement of the intrinsic angular momentum for an arbitrary state $|\psi\rangle$ in $z$ direction, $S_z$, is made, this state can be expressed by,

$$|\psi\rangle = c_+|+\mathbf{z}\rangle + c_-|-\mathbf{z}\rangle \tag{6}$$

where the arbitrary state $|\psi\rangle$ is in the superposition of the states $|+\mathbf{z}\rangle$ and $|-\mathbf{z}\rangle$ with amplitudes of $c_+$ and $c_-$, respectively. The amplitudes $c_+$ and $c_-$ depend on the orientation of the magnetic field that the particles pass through.

For a *ket vector* $|\psi\rangle$, a corresponding *bra vector* can be defined as, $\langle \psi|$. The amplitude of a particle in the spin state $|\psi\rangle$ to be in the state $|\varphi\rangle$ can be shown by $\langle \varphi|\psi\rangle$ (which is an inner product representation). Thus, for example, we can write,

$$\langle +\mathbf{z}|\psi\rangle = c_+\langle +\mathbf{z}|+\mathbf{z}\rangle + c_-\langle +\mathbf{z}|-\mathbf{z}\rangle = c_+ \tag{7}$$

$$\langle -\mathbf{z}|\psi\rangle = c_+\langle -\mathbf{z}|+\mathbf{z}\rangle + c_-\langle -\mathbf{z}|-\mathbf{z}\rangle = c_- \tag{8}$$

Substituting $c_+$ and $c_-$ from Equations (7) and (8) into Equation (6) we obtain,

$$|\psi\rangle = \langle +\mathbf{z}|\psi\rangle|+\mathbf{z}\rangle + \langle -\mathbf{z}|\psi\rangle|-\mathbf{z}\rangle \tag{9}$$

By placing the amplitudes $\langle +\mathbf{z}|\psi\rangle$, and $\langle -\mathbf{z}|\psi\rangle$ after the *kets* in Equation (9), we have,

$$|\psi\rangle = |+\mathbf{z}\rangle\langle +\mathbf{z}|\psi\rangle + |-\mathbf{z}\rangle\langle -\mathbf{z}|\psi\rangle \tag{10}$$

The amplitudes are complex numbers, and therefore placing them before or after the *kets* in Equations (9) and (10) does not make any difference. An analogy to presenting the amplitudes and the states of the quantum form in a classical mechanics vector form, such as a force vector $\mathbf{F}$ in a two-dimensional problem, is the force $\mathbf{F} = F_x\mathbf{i} + F_y\mathbf{j}$, which is the same as $\mathbf{F} = \mathbf{i}F_x + \mathbf{j}F_y$.

By definition, there exists a *bra* vector for each *ket*. In order to complete the quantum notation for the states, the *bra* vector is also represented similarly to the *ket* representation of the amplitudes and the states as follows:

$$\langle \psi| = c'_+\langle +\mathbf{z}| + c'_-\langle -\mathbf{z}| \tag{11}$$

If the same procedure for *kets* as in Equations (7) to (10), is carried out for the *bra* representation in Equation (11), we have,

$$\langle\psi|+\mathbf{z}\rangle = c'_+\langle+\mathbf{z}|+\mathbf{z}\rangle + c'_-\langle+\mathbf{z}|-\mathbf{z}\rangle = c'_+ \quad (12)$$

$$\langle\psi|-\mathbf{z}\rangle = c'_+\langle+\mathbf{z}|-\mathbf{z}\rangle + c'_-\langle-\mathbf{z}|-\mathbf{z}\rangle = c'_- \quad (13)$$

and substituting the results in (12) and (13) into (11) leads to,

$$\langle\psi| = \langle\psi|+\mathbf{z}\rangle\langle+\mathbf{z}| + \langle\psi|-\mathbf{z}\rangle\langle-\mathbf{z}| \quad (14)$$

In order to evaluate $\langle\psi|\psi\rangle$, Equations (9) and (14) are combined, noting that $\langle+\mathbf{z}|+\mathbf{z}\rangle = 1$, $\langle-\mathbf{z}|+\mathbf{z}\rangle = 0$, $\langle-\mathbf{z}|-\mathbf{z}\rangle = 1$, $\langle+\mathbf{z}|-\mathbf{z}\rangle = 0$

$$\langle\psi|\psi\rangle = \langle\psi|+\mathbf{z}\rangle\langle+\mathbf{z}|\psi\rangle + \langle\psi|-\mathbf{z}\rangle\langle-\mathbf{z}|\psi\rangle \quad (15)$$

The term, $\langle\psi|\psi\rangle$, which represents the amplitude of a particle in the spin state $\psi$ to be in the state $\psi$, is equal to one, by definition.

For a case that $\langle\psi|+\mathbf{z}\rangle$ implies $\langle+\mathbf{z}|\psi\rangle$, and $\langle\psi|-\mathbf{z}\rangle$ implies $\langle-\mathbf{z}|\psi\rangle$, which in fact is true by definition, we can write Equation (15) as,

$$\langle\psi|\psi\rangle = |\langle+\mathbf{z}|\psi\rangle|^2 + |\langle-\mathbf{z}|\psi\rangle|^2 = 1 \quad (16)$$

The term $|\langle+\mathbf{z}|\psi\rangle|^2$ is referred to the probability of a particle in the state $|\psi\rangle$ being in the state $|+\mathbf{z}\rangle$ when $S_z$ is measured. Similarly, the statement for $|\langle+\mathbf{z}|\psi\rangle|^2$ is: the probability of a particle in the state $|\psi\rangle$ being in the state $|-\mathbf{z}\rangle$ when $S_z$ is measured. Equation (16) is interpreted as *the sum of the probabilities of finding the particle in the state $|+\mathbf{z}\rangle$ or $|-\mathbf{z}\rangle$ is equal to one*. Quantum physics, therefore, unlike any classical mechanics expectation, states that the probability amplitudes of $\langle+\mathbf{z}|\psi\rangle$ and $\langle-\mathbf{z}|\psi\rangle$ can be nonzero simultaneously, where a particle can be in the superposition of the states $|+\mathbf{z}\rangle$ and $|-\mathbf{z}\rangle$. Thus, measuring intrinsic angular momentum can result in non-zero probabilities for both $S_z = \hbar/2$ and $S_z = -\hbar/2$ simultaneously. In classical mechanics, we only expect to obtain one value for the angular momentum of a particle at a time.

## 5 Rotation of Basis States

A spin state $|\psi\rangle$ of a spin-$\frac{1}{2}$ particle is developed when sending particles through a magnetic field oriented in an arbitrary direction. The amplitudes $\langle+\mathbf{z}|\psi\rangle$ and $\langle-\mathbf{z}|\psi\rangle$ give the projected values of $|\psi\rangle$ onto the states $|+\mathbf{z}\rangle$ and $|-\mathbf{z}\rangle$, respectively (Equation (9)). As in a classical case for force vector, for instance, the amplitudes and the projected values are used in representing the vector $\mathbf{F}$, in a quantum scenario the projected values or the amplitudes are complex numbers representing a state $|\psi\rangle$.

In classical mechanics, a vector, for instance, force $\mathbf{F}$ is shown as $\mathbf{F} = F_x\mathbf{i} + F_y\mathbf{j}$, or $\mathbf{F} \rightarrow (F_x, F_y)$. Similarly, a quantum representation of a *ket* can be given by a column vector as,

$$|\psi\rangle \xrightarrow[S_z \text{ basis}]{} \begin{pmatrix}\langle+\mathbf{z}|\psi\rangle \\ \langle-\mathbf{z}|\psi\rangle\end{pmatrix} = \begin{pmatrix}c_+ \\ c_-\end{pmatrix} \quad (17)$$

Now in this vector form, the state $|+\mathbf{z}\rangle$ for instance, can be given by,

$$|+\mathbf{z}\rangle \xrightarrow[S_z \text{ basis}]{} \begin{pmatrix}\langle+\mathbf{z}|+\mathbf{z}\rangle \\ \langle-\mathbf{z}|+\mathbf{z}\rangle\end{pmatrix} = \begin{pmatrix}1 \\ 0\end{pmatrix} \quad (18)$$

and the state $|-\mathbf{z}\rangle$ is represented by,

$$|-\mathbf{z}\rangle \xrightarrow[S_z \text{ basis}]{} \begin{pmatrix}\langle+\mathbf{z}|-\mathbf{z}\rangle \\ \langle-\mathbf{z}|-\mathbf{z}\rangle\end{pmatrix} = \begin{pmatrix}0 \\ 1\end{pmatrix} \quad (19)$$

If a particle in the state $|+\mathbf{x}\rangle$ is sent through a magnetic field in the $z$ direction, and measurements of $S_z$ gives probabilities of 50% for both $\hbar/2$ and $-\hbar/2$, we have,

$$|\langle+\mathbf{z}|+\mathbf{x}\rangle|^2 = \tfrac{1}{2} \quad (20)$$

$$|\langle-\mathbf{z}|+\mathbf{x}\rangle|^2 = \tfrac{1}{2} \quad (21)$$

which allows to express Equation (5) in the vector form as,

$$|+\mathbf{x}\rangle \xrightarrow[S_z \text{ basis}]{} \begin{pmatrix}\langle+\mathbf{z}|+\mathbf{x}\rangle \\ \langle-\mathbf{z}|+\mathbf{x}\rangle\end{pmatrix} = \frac{1}{\sqrt{2}}\begin{pmatrix}1 \\ 1\end{pmatrix} \quad (22)$$

In order to write Equation (15) in a vector/matrix form, and accommodate the *bra* vector for $\langle\psi|$, the relation can be written as,

$$\langle\psi|\psi\rangle = (\langle\psi|+\mathbf{z}\rangle, \langle\psi|-\mathbf{z}\rangle)\begin{pmatrix}\langle+\mathbf{z}|\psi\rangle \\ \langle-\mathbf{z}|\psi\rangle\end{pmatrix} \quad (23)$$

It can be seen that the appropriate vector form for a *bra* is a row vector to satisfy the matrix multiplication of the *ket* and *bra* as in (23), and thus,

$$\langle\psi| \xrightarrow[S_z \text{ basis}]{} (\langle\psi|+\mathbf{z}\rangle, \langle\psi|-\mathbf{z}\rangle) \quad (24)$$

A rotation operator can be used to represent the transformation of a *ket* from one state to another. For instance, if the magnetic moment of a spin-$\frac{1}{2}$ particle in the state $|+\mathbf{z}\rangle$ placed in a magnetic field is in the $x$ direction, the spin will rotate in the $x$-$z$ plane and as time progresses, the particle will be in the state $|+\mathbf{x}\rangle$. This rotation of the state $|+\mathbf{z}\rangle$ in the magnetic field pointed in the $x$ direction can be denoted by $\hat{R}$. The rotation can be given by,

$$|+\mathbf{x}\rangle = \hat{R}\left(\frac{\pi}{2}\mathbf{j}\right)|+\mathbf{z}\rangle \quad (25)$$

where $\mathbf{j}$ indicates the unit vector along the axis of rotation y. Therefore, operator $\hat{R}$ rotates $|+\mathbf{z}\rangle$ about the unit vector $\mathbf{j}$ with an angle of $\frac{\pi}{2}$. Such an operator rotates an arbitrary state of a spin-$\frac{1}{2}$ particle $|\psi\rangle$, when placed in a magnetic field in the $x$ direction, as given by,

$$\hat{R}\left(\frac{\pi}{2}\mathbf{j}\right)|\psi\rangle = \hat{R}\left(\frac{\pi}{2}\mathbf{j}\right)(c_+|+\mathbf{z}\rangle + c_-|-\mathbf{z}\rangle) \quad (26)$$

$$= c_+\hat{R}\left(\frac{\pi}{2}\mathbf{j}\right)|+\mathbf{z}\rangle$$

$$+ c_-\hat{R}\left(\frac{\pi}{2}\mathbf{j}\right)|-\mathbf{z}\rangle$$

$$= c_+|+\mathbf{x}\rangle + c_-|-\mathbf{x}\rangle$$

In Equation (26), it is assumed that the state $|\psi\rangle$ is initially in superposition of the states $|+\mathbf{z}\rangle$ and $|-\mathbf{z}\rangle$ as in Equation (6). Note that the amplitudes $c_+$ and $c_-$ are numbers, which can be placed before or after the operator.

An infinitesimal rotation, for instance, by an angle $d\phi$ about the $z$ axis can be expressed by an operator $\hat{R}(d\phi\mathbf{k})$ given by,

$$\hat{R}(d\phi\mathbf{k}) = 1 - \frac{i}{\hbar}\hat{J}_z d\phi \quad (27)$$

$\hat{J}_z$ is an operator that generates rotations about the $z$ axis. As seen from (25), when $d\phi \to 0$, $\hat{R}(d\phi\mathbf{k}) \to 1$. $\hat{J}_z$ have the same dimensions of as $\hbar$, or angular momentum.

# 6 The Schrödinger Equation

Time evolution in quantum mechanics is introduced by the Hamiltonian operator, which gives the time translations in quantum systems. The time-evolution operator $\hat{U}(t)$ translates a *ket* vector forward in time as given by,

$$\hat{U}(t)|\psi(0)\rangle = |\psi(t)\rangle \quad (28)$$

where $|\psi(t)\rangle$ is the state of the system at time $t$, with the initial state $|\psi(0)\rangle$ at time $t=0$.

By adopting the rotation operator in (27), a time translation, as an operator, and knowing that $|\psi(t)\rangle$ is a unit vector, $\hat{U}(t)$ must be a unitary transformation and therefore, the first-order Taylor expansion of $\hat{U}(t)$ can be given by,

$$\hat{U}(t) \approx 1 - \frac{i}{\hbar}\hat{H}dt \quad (29)$$

where the operator $\hat{H}$ is the generator of time translations. The operator $\hat{H}$ translates an initial *ket* to a *ket* at a different time. By substituting (29) into (28) we get,

$$|\psi(t)\rangle = (1 - \frac{i}{\hbar}\hat{H}dt)|\psi(0)\rangle \quad (30)$$

Equation (30) can be rewritten as,

$$|\psi(t)\rangle - |\psi(0)\rangle = \left(-\frac{i}{\hbar}\hat{H}dt\right)|\psi(t)\rangle \quad (31)$$

Or

$$i\hbar\frac{d}{dt}|\psi(t)\rangle = \hat{H}|\psi(t)\rangle \quad (32)$$

Equation (32) is known as the Schrödinger Equation. This equation describes the state of the system at its position at a corresponding time. This is analogous to Newton's second law of motion, which describes the state of a physical system at each time instant by solving for the position and momentum of the system as a function of the applied forces.

# 7 Controls in the Context of Quantum States

Dynamics of a robotic system, and of a robotic control task, as an example of a control of multi-agent systems in a network of an autonomous system, has been presented in Section 3.

The dynamics of a robotic platform can be given by Equations (1) and (2). For a multi-agent networked system, the equations of motion of the agents can be given by the system of equations, which in a matrix form can be represented by $\{\mathbf{F}\} = [\mathbf{M}]\{\mathbf{a}\}$. The present paper aims to integrate the classical dynamics problems with quantum capabilities such as quantum entanglement and cryptography, to enhance the performance of dynamic systems, through an interdisciplinary approach. The integrated quantum and classical control block diagram of a dynamic system is presented in Fig. 3.

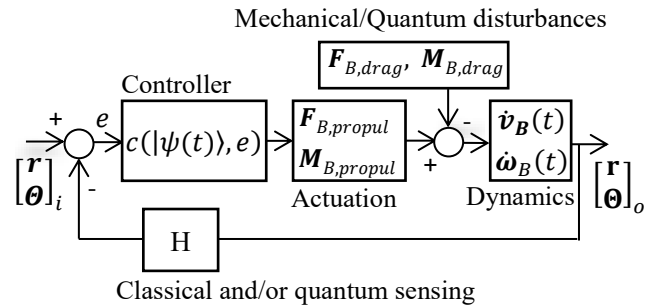

**Fig. 3. The control block diagram for an autonomous dynamic system with an integrated quantum control system.**

The Laplace transform of the integrated classical and quantum feedback control system in Fig. 3 is of interest. The Laplace transform approach for classical dynamic systems is readily available. The Laplace transform approach for quantum states is presented below, and further, it is proposed for the integrated control system. The Schrödinger Equation in (32) is repeated by,

$$i\hbar\frac{d}{dt}|\psi(t)\rangle = \hat{H}|\psi(t)\rangle$$

The Hamiltonian in the Schrödinger Equation can be expressed by $\hat{H} = \hat{H}^0 + \hat{H}'$, where $\hat{H}'$ is the perturbation.

By applying the Laplace transformation to the Schrödinger Equation [53], we get,

$$\mathcal{L}[\hat{H}|\psi(t)\rangle] = \mathcal{L}\left[i\hbar \frac{d}{dt}|\psi(t)\rangle\right] \quad (33)$$

This can be written as [53],

$$\hat{H}^0 \mathcal{L}[|\psi(t)\rangle] + \mathcal{L}[\hat{H}'|\psi(t)\rangle] = -i\hbar|\psi(0)\rangle + i\hbar\mathcal{L}\left[\frac{d}{dt}|\psi(t)\rangle\right] \quad (34)$$

Therefore,

$$\hat{H}^0|\Psi(s)\rangle + \mathcal{L}[\hat{H}'|\psi(t)\rangle] = -i\hbar|\psi(0)\rangle + i\hbar s|\Psi(s)\rangle \quad (35)$$

where $\mathcal{L}[|\psi(t)\rangle]$ is denoted by $|\Psi(s)\rangle$.

The time-independent and time-dependent cases of $\hat{H}'$ are presented now.

In the first case, where $\hat{H}'$ is time-independent, the solution to the expression in (35) can be given by,

$$(\hat{H}^0 + \hat{H}' - i\hbar s)|\Psi(s)\rangle = -i\hbar|\psi(0)\rangle \quad (36)$$

Equation (36) represents linear variation of a system with the generalized inertia matrix (Laplace transform of the system matrix),

$$M_l = \langle\Psi(s)|\hat{H} - i\hbar s|\Psi(s)\rangle + i\hbar\langle\Psi(s)|\psi(0)\rangle + i\hbar\langle\psi(0)|\Psi(s)\rangle \quad (37)$$

Equation (34) is obtained from the extremals of $|M_l|^2$.

Inverse Laplace transform of the time-independent case can be obtained by rewriting Equation (36) as,

$$(\hat{H}^0 + \lambda\hat{H}' - i\hbar s)|\Psi(s)\rangle = -i\hbar|\psi(0)\rangle \quad (38)$$

where $\lambda$ is a perturbation parameter, and by expanding $|\Psi(s)\rangle$ in $\lambda^n$, as,

$$|\Psi(s)\rangle = \sum_{n=0} \lambda^n |\Psi^n(s)\rangle \quad (39)$$

By substituting (39) into (46) and separating the terms of each order in $\lambda$ we have,

$$(\hat{H}^0 - i\hbar s)|\Psi^{(0)}(s)\rangle = -i\hbar|\psi(0)\rangle$$
$$(\hat{H}^0 - i\hbar s)|\Psi^{(1)}(s)\rangle + \hat{H}'|\Psi^{(0)}(s)\rangle = 0 \quad (40)$$
$$(\hat{H}^0 - i\hbar s)|\Psi^{(2)}(s)\rangle + \hat{H}'|\Psi^{(1)}(s)\rangle = 0$$

For initial condition $|\psi(0)\rangle = \varphi_1^0$, the coupled equations in (40) are solved by expanding $|\Psi^n(s)\rangle$ in terms of eigenfunctions of $\hat{H}^0$ as [55],

$$|\Psi^n(s)\rangle = \sum_m C_m^{(n)} s\, \varphi_1^0 \quad (41)$$

By substituting (41) into (40), we get,

$$C_1^{(1)} = \frac{i\hbar}{i\hbar s - E_1^0} \quad (42)$$
$$C_1^{(0)} = 0$$

and also,

$$C_m^{(1)} = \frac{i\hbar\langle\varphi_m^0|\hat{H}'|\varphi_1^0\rangle}{(i\hbar s - E_m^0)(i\hbar s - E_1^0)} \quad (43)$$

which gives the first-order approximation as,

$$|\Psi(s)\rangle = \frac{i\hbar\varphi_1^0}{i\hbar s - E_1^0} + \lambda \sum_m \frac{i\hbar\langle\varphi_m^0|\hat{H}'|\varphi_1^0\rangle \varphi_m^0}{(i\hbar s - E_m^0)(i\hbar s - E_1^0)} + \cdots \quad (44)$$

Finally, the inverse Laplace transform is determined from (44), as,

$$|\psi(t)\rangle = \left[1 - \frac{it\lambda}{\hbar\langle\varphi_1^0|\hat{H}'|\varphi_1^0\rangle}\right] \varphi_1^0 \exp\left(-\frac{iE_1^0 t}{\hbar}\right) + \lambda \sum_m \frac{i\hbar\langle\varphi_m^0|\hat{H}'|\varphi_1^0\rangle \varphi_m^0}{(E_m^0 - E_1^0)} \left\{\exp\left(-\frac{iE_1^0 t}{\hbar}\right) - \exp\left(-\frac{iE_1^0 t}{\hbar}\right)\right\} \varphi_m^0 \quad (45)$$

For the second case, $\hat{H}'$ is time-dependent and given by [54],

$$\hat{H}' = \sum_l V_l(r) exp(-il\epsilon t/\hbar) \quad (46)$$

where $l$ ranges from $-\infty$ to $\infty$.

By applying the Laplace transformation to the (32), we get,

$$\hat{H}^0|\Psi(s)\rangle + \sum_l V_l(r)\mathcal{L}[exp(-il\epsilon t/\hbar)|\psi(t)\rangle] = -i\hbar|\psi(0)\rangle + i\hbar s|\Psi(s)\rangle \quad (47)$$

By Taylor's expansion of the second term in (47) is expressed as,

$$\hat{H}^0|\Psi(s)\rangle + \sum_l V_l(r) exp(-il\epsilon D_s/\hbar)\, |\Psi(s)\rangle = -i\hbar|\psi(0)\rangle + i\hbar s|\Psi(s)\rangle \quad (48)$$

where $D_s$ denotes the deferential operator $\frac{\partial}{\partial s}$. Equation (48) can be rewritten as,

$$\left[\hat{H}^0 + \sum_l V_l(r) exp(il\epsilon D_s/\hbar) - i\hbar s\right]|\Psi(s)\rangle = -i\hbar|\psi(0)\rangle \quad (49)$$

which is the linear variation of

$$N_l = \langle\Psi(s)|\hat{H}^0 + + \sum_l V_l(r) exp(il\epsilon D_s/\hbar) - i\hbar s|\Psi(s)\rangle + i\hbar\langle\Psi(s)|\psi(0)\rangle + i\hbar\langle\psi(0)|\Psi(s)\rangle \quad (50)$$

The extremals of $|N_l|^2$ obtain Equation (49).

The transfer function of the feedback control system in Fig. 3 may be expressed as,

$$TF = c(|\Psi(s)\rangle, e)Act(\{\mathbf{F}(s), |\Psi(s)\rangle\})Dyn(\{\mathbf{F}(s)\}) \quad (51)$$
$$/1 + c(|\Psi(s)\rangle, e)Act(\{\mathbf{F}(s)\}, |\Psi(s)\rangle))Dyn(\{\mathbf{F}(s)\})\mathbf{H}$$

The transfer function in (51) is a generalized feedback control system, which integrates the classical dynamics of the system, actuation, and sensing, denoted by $Dyn(\{\mathbf{F}(s)\})$, $Act(\{\mathbf{F}(s)\})$, and $\mathbf{H}$, respectively, with the quantum states $|\Psi(s)\rangle$. The response in the time domain is obtained by applying a desired input to the transfer function and eventually through the inverse Laplace of the function.

# 8 Quantum Cryptography for Automation

An automation system that includes robotic manipulators, and an instrumented conveyor with integrated sensors and actuators, and autonomous mobile platforms, is shown in Fig. 4. In this section, it is desired to implement automated control tasks in such automation scenarios by using quantum cryptography techniques. A generalized feedback control system has been introduced in Section 7. Quantum cryptography is applied to guarantee security against cyber-physical attacks while performing efficient control tasks for autonomy. A schematic diagram of a quantum cryptography process is given in Fig. 5. Fig. 6 and Fig. 7 show the components of the robotic systems in more detail. Fig. 8 introduces an alternative mobile platform as an aerial system rather than a ground platform.

Experimental quantum cryptography is carried out by polarizing photons, passing them through a polarizing beamsplitter cube, and detecting the photons' polarization.

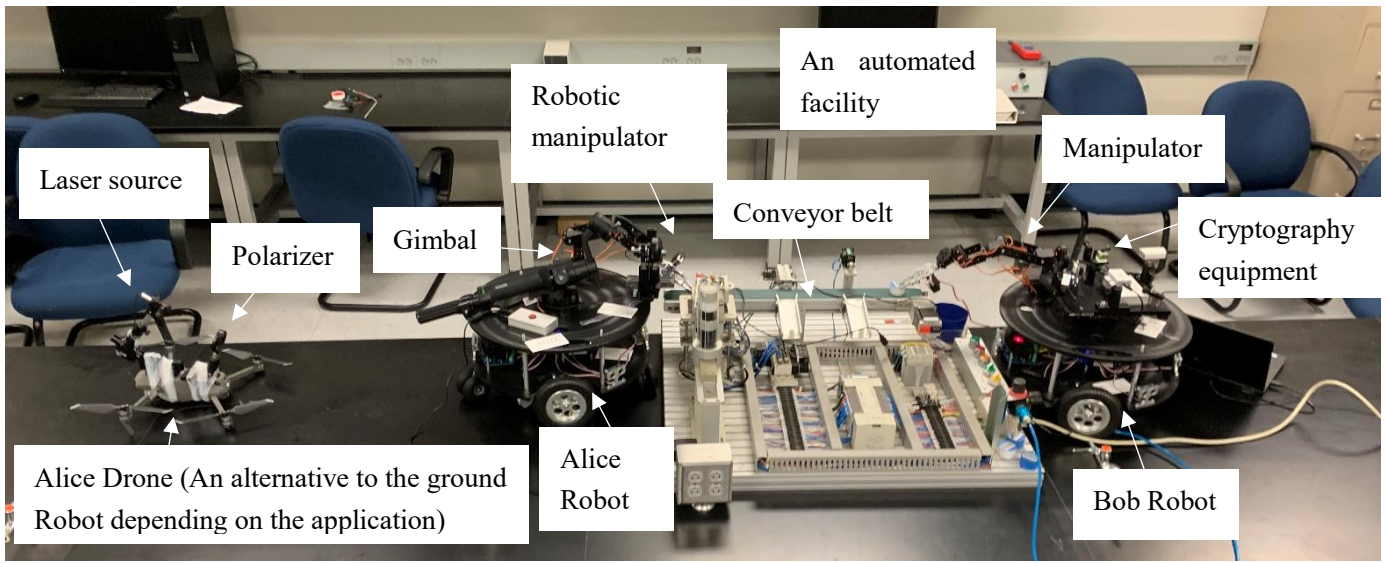

**Fig. 4 An integrated automation and robotic system with quantum cryptography instrumentation.**

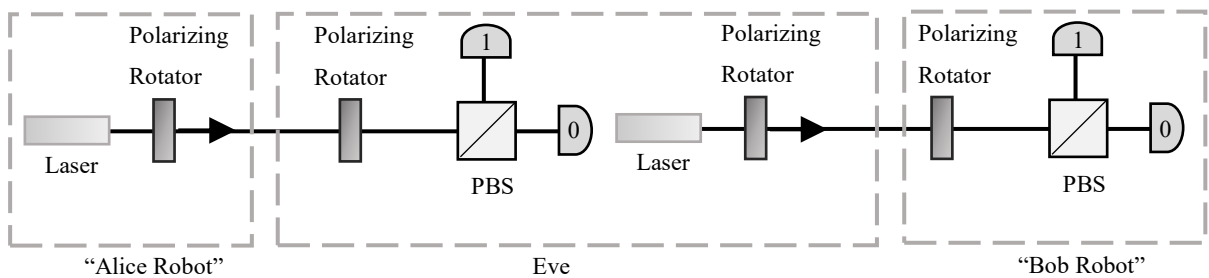

**Fig. 5 Quantum cryptography experimental setup.**

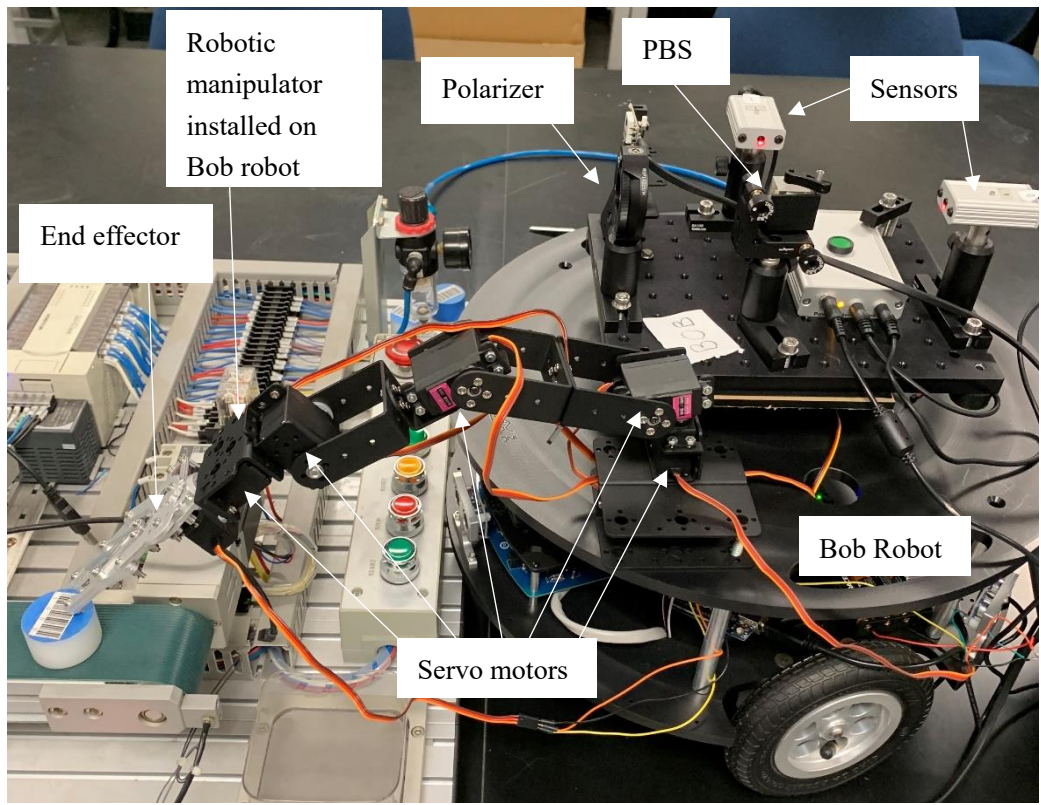

**Fig. 6 The Bob Robot.**

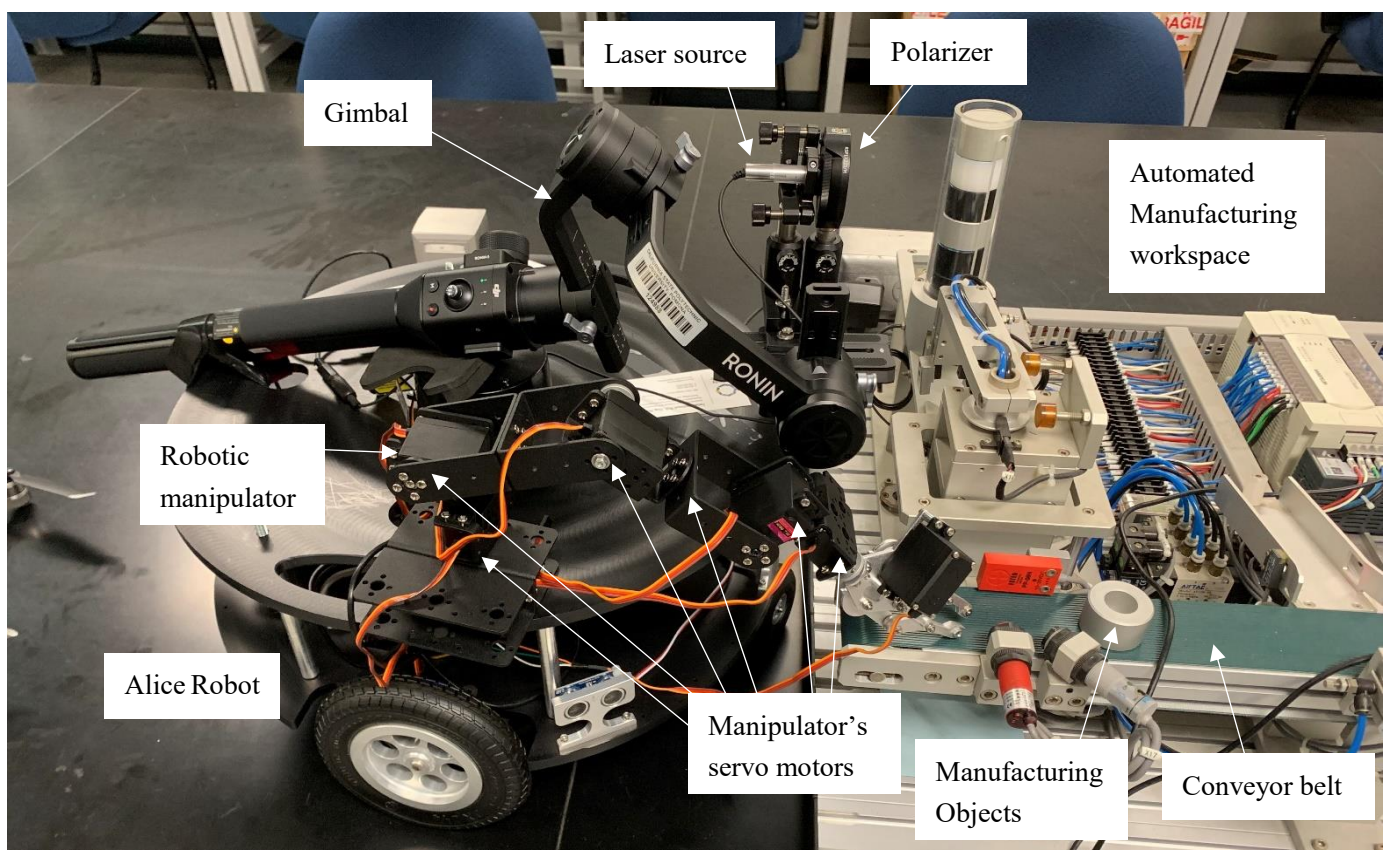

**Fig. 7 The Alice Robot.**

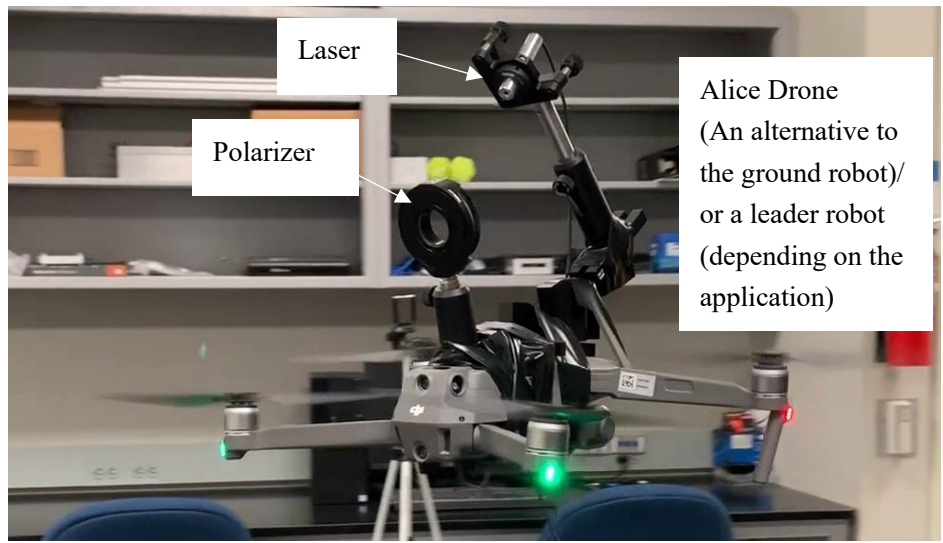

**Fig. 8 The Alice drone.**

The primary operations in a quantum cryptography experiment include:
- Generating and sending single photons by a laser diode.
- Spontaneous Parametric Down-Conversion (SPDC) process.
- Polarizing photons: performed by sending photons through $\frac{\lambda}{2}$ (half-wave) plate.
- The process of detecting the polarizations: the polarized photons are sent to a polarizing beamsplitter (PBS) cube, where the PBS passes the horizontal polarizations and reflects vertical polarizations, and two sensors, each dedicated to vertical or horizontal polarization sensing, respond to the receiving of the photons.

For the laboratory demonstration of experimental quantum cryptography, we assume that the laser diode is producing single photons, by applying the SPDC process, and the sensors detect single photons. For demonstration purposes, instead of single photons, laser pulses are used, which is sufficient for the experimental demonstration of the concept of experimental quantum cryptography here. The process of generating and counting (detecting) single photons is presented in the entanglement experiment in Section 9. The mobile robots in conjunction with quantum cryptography tools are used for cooperative tasks and control of the mobile robot applications. The "Alice robot" contains a laser diode and a $\frac{\lambda}{2}$ polarizing rotator plate. The "Bob robot" contains a $\frac{\lambda}{2}$ polarizing rotator plate, a polarizing beamsplitter cube, and two sensors. Eve, eavesdrop attacker, tries to: a) intercept and detect the information sent from Alice, and b) duplicate the information and send it to Bob. In the quantum cryptography, Eve is exposed to Alice and Bob after a few exchanges of photons, because a single photon cannot be partially detected (as once a single photon is detected by a sensor, it is trapped). Also, the polarization of a single photon (e.g., vertical and horizontal) can be produced by a variety of polarization combinations. When a photon passes through two separate polarizers, one on Alice and the other on Bob, the result of two polarizers are combined. So, by having two polarizers (one on Alice, and the other on Bob), there are various ways of combining polarizations to achieve one result. This is discussed next.

The "ket" symbol $|\ \rangle$ is used to denote a quantum state. The combination of polarization associated with the half-wave plates at Alice and Bob are given in [49]-[50], where "+" basis corresponds to $|0º\rangle$, and $|90º\rangle$, polarizations (or horizontal and vertical polarizations) and "×" basis corresponds to $|-45º\rangle$ and $|45º\rangle$ polarizations produced by the half-wave plate polarizers. The polarization states of the photons (i.e., $|-45º\rangle$, $|0º\rangle$, $|45º\rangle$, and $|90º\rangle$) from Alice to Bob are not known publicly. However, Alice and Bob publicly share the information about the bases ("+" or "×") they used. More details on the quantum cryptography procedure are found in [50].

The polarization of the photons reaching the detectors on the Bob Robot is converted to zeros and ones, where zeros and ones are associated with the horizontal and vertical polarizations, respectively. The zeros and ones are sent to a microcontroller, as digital commands, for control and autonomy applications.

# 9 Quantum Entanglement for Automation

Quantum entanglement can be achieved by the Spontaneous Parametric Down-Conversion (SPDC) process, where two particles are entangled, and predicts non-local behavior. Photons can be the states of vertical or horizontal polarization, and the entangled photons can be specified by the superposition of two photons that are in orthonormal linear polarizations (vertical and horizontal). Quantum mechanics only predicts that the photons are in the vertical and horizontal polarization states, simultaneously, but the state of the polarization of the photons cannot be individually labeled for each photon. When a measurement is made, we can find each of the two photons in one of two states with a corresponding probability. By knowing the polarization of one photon, the polarization of the other photon can be predicted as the polarizations of the photons are orthogonal in the SPDC process. However, the two photons are only in the entangled state until the measurement is made. Once the measurement is made, the photons are no longer correlated, and therefore they are no longer entangled, due to the non-local property of quantum mechanics (Violation of Bell's inequalities [63]). Non-local properties have led to many remarkable applications such as quantum teleportation (e.g., [64]-[68]), and the rise of the new field of quantum information (e.g., [69]-[71]).

A quantum entanglement experiment is shown in the setup of Fig. 9 (in this figure, the notations are: FM: Flipper mirror; M: Mirror; MA: Mirror A; MB: Mirror B; HWP: Half waveplate; AP: Autonomous (Mobile) Platform). The physical experimental setup is presented in Fig. 10. The SPDC process converts one photon of higher energy into a correlated pair of photons with lower energy. By sending a laser beam (e.g., a 100 mW laser, with 405 nm wavelength) through a nonlinear crystal (i.e., BBO: beta Barium Borate), one photon of higher energy is converted into a correlated photon pair (with 810nm wavelengths), producing entangled states, with orthogonal horizontal and vertical polarizations. The two single-photon counter module modules (SPCM50A/M [59]), in Fig. 10, installed on two robots, count photons (by detecting the incident photon), which is used for evaluating entanglement. The SPCM50A/M module specifications include: Wavelength Range of 350 - 900 nm, Typical Max Responsivity of 35% at 500 nm, and Active Detector Size 50 μm ([59]). The 810 nm narrow band-pass filters, with the Bandwidth of 30 nm, shown in Fig. 10, will allow only 810 nm photon pairs produced by the SPDC process to reach the Single Photon Counter Modules (SPCMs).

The process of identifying the entangled correlated photon pairs entails the detection of the photon pairs that reach the two SPCM detectors at the same time [74]. The signals from the two SPCM detectors can be sent to an electronic coincidence unit. Each SPCM detector assigns a coincidence to any pair of detected pulses that arrive within a specified time (basically an AND gate). Another possibility could be to have a circuit that assigns a time to the arrival of the photons so that through a classical channel, Alice and Bob can compare the arrival times, and those that arrive within a specified time can be considered in as coincidences. The SPCM detectors on Alice and Bob should synchronize their clocks to within tens of nanoseconds or perhaps pick the time from a wireless signal, and then record the time of arrival of the detector photons. Alice and Bob need to share the basis information through the classical channel (e.g., BB84 encoding), and also the photon arrival times [74].

If the HWP and the polarizer in Fig. 9 and Fig. 10 are used to manipulate the polarization state of one entangled photon (e.g., Alice), along the way when the photon travels from the BBO to the polarizing beamsplitter, the polarization of the other corresponding entangled photon (Bob) is affected simultaneously due to the entanglement phenomenon; this retains the photons in an entangled state (until before the measurement of a photon is made). This is where the non-classical phenomenon of the quantum entanglement appears when Alice and Bob are at any arbitrary distance apart.

When the single-photon counters detect entangled photon pairs, a digital signal from the counters can be sent to the robot onboard microcontroller, which then can be translated to a digital task for control of the autonomous system (such as a system of cooperating robots). It should be noted that although the wave functions of the entangled photons collapse after the detections, by converting the detected entangled photons to digital signals, the signals are now available for digital control purposes of the autonomous system. Once the Alice and Bob robots are in an entangled state, any desired information can be exchanged (e.g., by the quantum cryptography process) between a leader robot and the entangled robots for the two robots (Fig. 9 and Fig. 10) to perform the desired robotic or autonomous tasks.

The quantum cooperative autonomous platforms presented in the cryptography and entanglement section of the present paper, and a combination of scenario of entanglement and cryptography for autonomy, (where entanglement triggers the process of Cryptography for multiple robotic systems) perhaps, can be the most sophisticated technique in cooperative robotics and unmanned systems technology. This is due to the ultimate speed of photon propagation for robotic control applications, true guaranteed security and immunity against cyberattacks, and the possibility of having access to the entanglement capabilities (that does not exist in the classical domain).

An alternative way of quantum communication between autonomous systems is Quantum Teleportation (e.g., [64]-[68]). A brief description of this technique is given now. Assume a two-robot problem where Alice robot is to teleport a quantum state, for example $|\psi\rangle$, to Bob. Alice and Bob share a quantum entangled state via SPDC process (Fig. 9 and Fig. 10). Alice, with the state $|\psi\rangle$ to teleport, carries out a measurement on the entangled photon that is received from the SPDC process. Alice then communicates the outcome of the measurement classically to Bob. This is done by converting Bell States (e.g., horizontal and vertical) to corresponding digital codes (e.g., zero and one) and transferring the zero and one through a classical channel. Finally, Bob robot performs a single-qubit operation based on this information and retrieves the initial unknown state $|\psi\rangle$ [75]. It should be noted that in this Quantum Teleportation technique, the state of $|\psi\rangle$ is teleported from Alice to Bob, where the intermediate entanglement process that sends a pair of entangled photons (e.g., with vertical and horizontal polarizations) to Alice and Bob allows this quantum state teleportation process.

As supplemental material, some videos of the quantum robotic experiments, presented in this paper, are available in References [72] and [73].

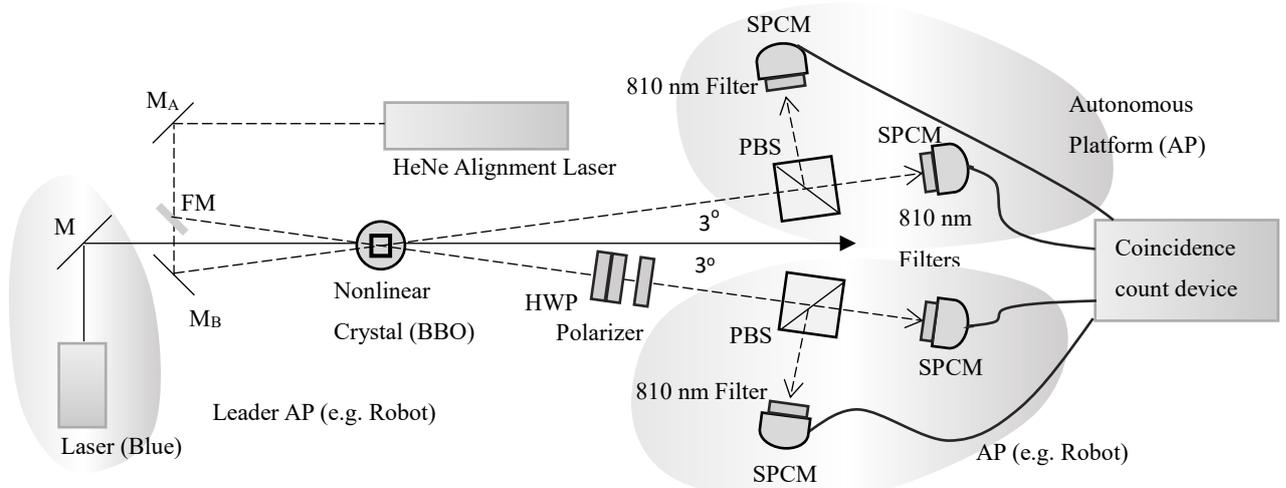

**Fig. 9 Quantum entanglement experimental setup diagram.**

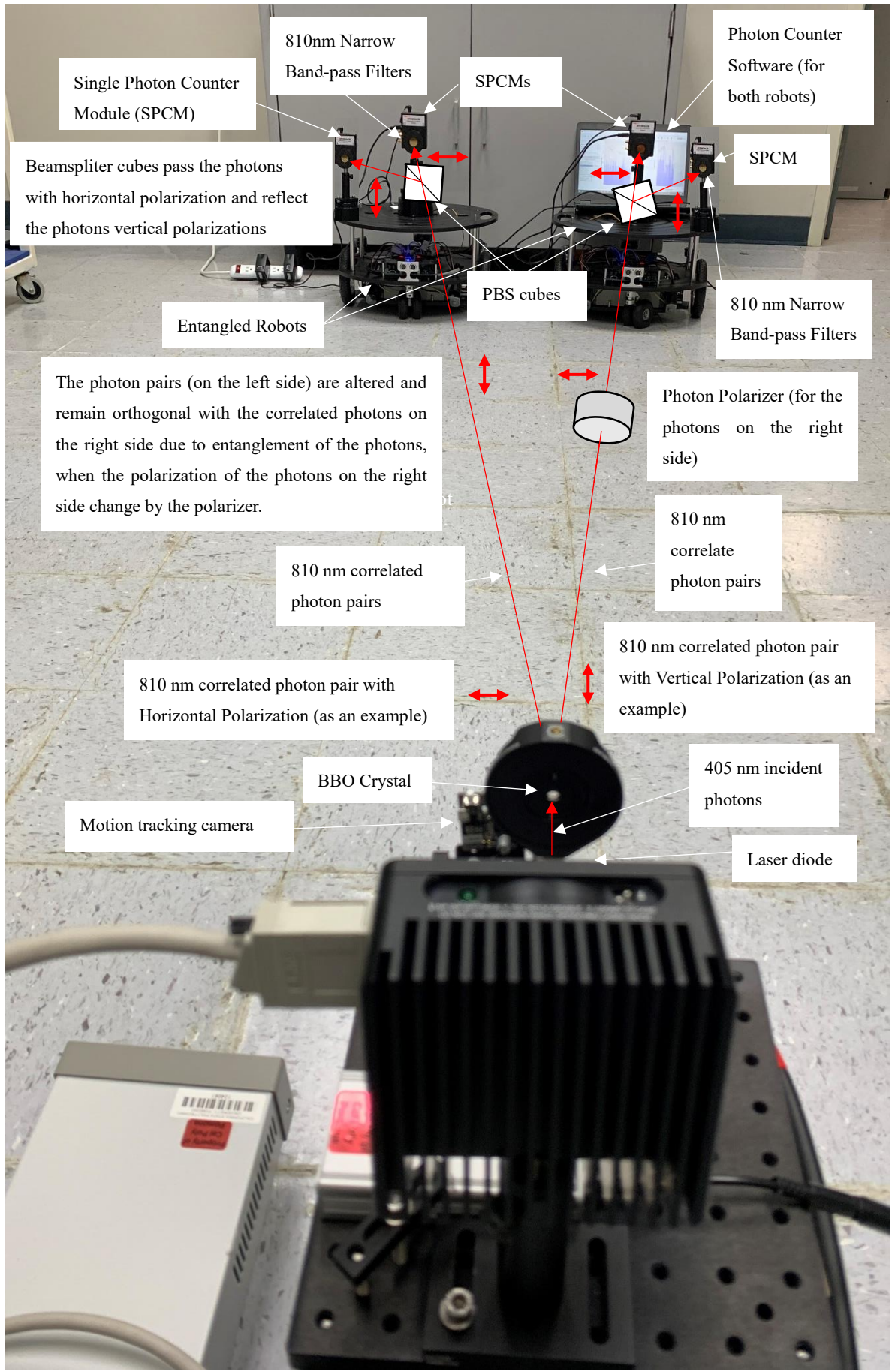

**Fig. 10 Quantum Entangled robot experimental setup.**

# 10 Conclusions

An integrated quantum and classical control system for autonomy was introduced in this paper. The dynamics and control system for a generalized automation scenario was presented as an application example. An introduction to analyzing the quantum states was given. A transfer function for the integrated quantum and classical feedback control system of the automated scheme was proposed. Quantum cryptography and entanglement were explored with applications to autonomous systems, and the corresponding experimental setups and procedures were described. The present paper proposed and discussed the integrated quantum and classical feedback control transfer function for the autonomy of mechanical systems, at non-atomistic scales, for the first time.

The future work of this research will include presenting and investigating of the use of quantum teleportation for control of autonomous and dynamical systems.

# References

[1] Feynman, R. P. (1982). Simulating physics with computers. *International Journal of Theoretical Physics*. 21 (6), pp. 467–488, 1982.

[2] Benioff, P. (1980). The computer as a physical system: A microscopic quantum mechanical Hamiltonian model of computers as represented by Turing machines. *Journal of Statistical Physics*, 22 (5), pp. 563-591.

[3] Paul Benioff, P. (1998). Quantum robots and environments, *Physical Review A*, Volume 58, Number 2, pp. 893-904.

[4] Paparo, G. D., Dunjko, V., Makmal, A., Martin-Delgado, M. A., and H. J. Briegel. (2014). Quantum Speedup for Active Learning Agents. *Physical Review X*, 4, 031002.

[5] Aiello, C. D., Cappellaro, P. (2015). Time-optimal control by a quantum actuator," *Physical Review A*, 91, 042340.

[6] Ajoy, A., Bissbort, U., Lukin, M. D., Walsworth, R. L., Cappellaro, P. (2015). Atomic-scale nuclear spin imaging using quantum-assisted sensors in diamond. *Physical Review X*, 5, 011001.

[7] A. Ajoy, A., Cappellaro, P. (2012). Stable three-axis nuclear-spin gyroscope in diamond. *Physical Review A*, 86, 062104.

[8] Ajoy, A., Liu, Y.-X., Saha, K., Marseglia, L., Jaskula, J.-C., Bissbort, U., Cappellaro, P. (2017). Quantum interpolation for high-resolution sensing. *Proceedings of the National Academy of Sciences*, 114 (9) 2149-2153; DOI: 10.1073/pnas.1610835114

[9] Cooper, A., Magesan, E., Yum, H. N., Cappellaro, P. (2014). Time-resolved magnetic sensing with electronic spins in diamond. *Nature communications*, 5:3141. doi: 10.1038.

[10] Cappellaro, P., Emerson, J., Boulant, N., Ramanathan, C., Lloyd, S., Cory, D. G. (2005). Entanglement Assisted Metrology. *Physical Review Letters*, 94, 020502.

[11] Goldstein, G., Cappellaro, P., Maze, J. R., Hodges, J. S., Jiang, L., Sørensen, A. S., and Lukin, M. D. (2011). Environment-Assisted Precision Measurement. *Physical Review Letters*, 106, 140502.

[12] Hirose, M., Aiello, C. D., and Cappellaro, P. (2012). Continuous dynamical decoupling magnetometry," *Physical Review A*, 062320.

[13] Magesan, E., Cooper, A., Yum, H., Cappellaro, P. (2013). Reconstructing the profile of time-varying magnetic fields with quantum sensors. *Physical Review A*, 88, 032107.

[14] Maze, J. R., Stanwix, P. L., Hodges, J. S., Hong, S., Taylor, J. M., Cappellaro, P., Jiang, L., Gurudev Dutt, M. V., Togan, E., Zibrov, A. S., Yacoby, A., Walsworth, R. L., Lukin, M. D. (2008). Nanoscale magnetic sensing with an individual electronic spin in diamond. Nature volume 455, pp. 644-647. doi:10.1038/nature07279.

[15] Taylor, J. M., Cappellaro, P., Childress, L., Jiang, L., Budker, D., Hemmer, P. R., Yacoby, A., Walsworth, R., Lukin, M. D. (2008). Highsensitivity diamond magnetometer with nanoscale resolution," *Nature Physics*, Volume 4, pp. 810–816.

[16] Joseph Kerckhoff, J. (2011). Quantum Engineering with Quantum Optics. Ph.D. thesis, Stanford University, CA, USA.

[17] Mabuchi H., Khaneja, N. (2005). Principles and applications of control in quantum systems. *Int. J. Robust Nonlinear Control* 15, 647-667.

[18] Mabuchi, H. (2009). Continuous quantum error correction as classical hybrid control. *New Journal of Physics*, Volume 11, 105044.

[19] Doherty, A. C., Doyle, J. C., Mabuchi, H., Jacobs, K., Habib, S. (2000). Robust control in the quantum domain. *Proceedings of the 39th IEEE Conference on Decision and Control*, Number 6939802, Sydney, NSW, Australia. DOI: 10.1109/CDC.2000.912895

[20] Doherty, A. C., Habib, S., Jacobs, K., Mabuchi, H., Tan, S. M. (2000). Quantum feedback control and classical control theory. *Phys. Rev. A*, 62, 012105.

[21] Mabuchi, H. (2008). Coherent-feedback quantum control with a dynamic compensator. *Physical Review A*, 78, 032323.

[22] Mabuchi, H. (2011). Coherent-feedback strategy to suppress spontaneous switching in ultra-low power optical bistability.

*Appl. Phys. Lett. 98*, 193109.

[23] Stockton, J., Armen, M., Mabuchi, H. (2002). Programmable logic devices in experimental quantum optics. *Journal of the Optical Society of America B*, Vol. 19, Issue 12, pp. 3019-3027. doi:10.1364/JOSAB.19.003019.

[24] Ghafari, F., Tischler, N., Di Franco, C., Thompson, J., Gu, M., Pryde, G. J. (2019). Interfering trajectories in experimental quantum-enhanced stochastic simulation, Nature Communications, 10:1630.
https://doi.org/10.1038/s41467-019-08951-2

[25] Arrangoiz-Arriola, P., Wollack, E. A., Wang, Z., Pechal, M., Jiang, W., McKenna, T. P., Witmer, J. D., Van Laer, R., Safavi-Naeini, A. (2019). Resolving the energy levels of a nanomechanical oscillator. *Nature Letter*, Vol 571, pp. 537 -547.
https://doi.org/10.1038/s41586-019-1386-x

[26] Li, L., Zhang, R., Zhao, Z., Xie, G., Liao, P., Pang, K., Song, H., Liu, C., Ren, Y., Labroille, G., Jian, P., Starodubov, D., Lynn, B., Bock, R., Tur, M., Willner, A. E. (2017). High-Capacity Free-Space Optical Communications Between a Ground Transmitter and a Ground Receiver via a UAV Using Multiplexing of Multiple Orbital-Angular-Momentum Beams, *Scientific Reports*, volume 7, Article number: 17427.

[27] Casimir, H. B. G. (1932). Rotation of a Rigid Body in Quantum-mechanics, *Nature*, volume 129, page780.
https://doi.org/10.1038/129780b0DO

[28] Khoshnoud, F., Quadrelli, M. B., Esat, I. I., de Silva, C. W., Quantum Multibody Dynamics, 2020, in progress.

[29] Song, S., Hayashi, M. (2018). Secure Quantum Network Code without Classical Communication. *IEEE Information Theory Workshop*, Guangzhou, China, DOI: 10.1109/ITW.2018.8613516.

[30] Tittel, W., Brendel, J., Gisin, B., Herzog, T., Zbinden, H., Gisin, N. (1998). Experimental demonstration of quantum correlations over more than 10 km, *Physical Review A*, Vol. 57, No. 5, pp. 3229-3232.

[31] Schumacher, B. (1996). Sending entanglement through noisy quantum channels. *Physical review A*. 54(4): 2614-2628.

[32] Ursin, R., Tiefenbacher, F., Schmitt-Manderbach, T., Weier, H., Scheidl, T., Lindenthal, M., Blauensteiner, B., Jennewein, T., Perdigues, J., Trojek, P., Ömer, B., Fürst, M., Meyenburg, M., Rarity, J., Sodnik, Z., Barbieri, C., Weinfurter, H., Zeilinger, A., (2007). Entanglement-based quantum communication over 144 km. *Nature Physics*, volume 3, pages 481–486.

[33] Resch, K. J., Lindenthal, M., Blauensteiner, B., Böhm, H. R., Fedrizzi, A., Kurtsiefer, C., Poppe, A., Schmitt-Manderbach, T., Taraba, M., Ursin, R., Walther, P., Weier, H., Weinfurter, H., Zeilinger, A. (2005). Distributing entanglement and single photons through an intra-city, free-space quantum channel. *Optics Express*, No. 1. Vol. 13, Issue 1, pp. 202-209.
DIO: 10.1364/OPEX.13.000202

[34] Waks, E., Zeevi, A., Yamamoto, Y. (2002). Security of quantum key distribution with entangled photons against individual attacks. *Physical Review A*, Vol. 65, 052310.

[35] Summhammer, J. (2006). Quantum Cooperation of Two Insects. arXiv:quant-ph/0503136.

[36] Iimura, Takeda K., Nakayama, S. (2015). Effect of Quantum Cooperation in Three Entangled Ants, *Int. J. of Emerging Technology and Advanced Engineering*, Vol. 5, Issue 8, pp. 29-33.

[37] Khoshnoud, F., de Silva, C. W., Esat, I. I. (2015). Bioinspired Psi Intelligent control for autonomous dynamic systems. *Journal of Control and Intelligent Systems*, Vol. 43, No. 4., pp. 205-211.

[38] Khoshnoud, F., de Silva, C. W., Esat, I. I. (2017). Quantum Entanglement of Autonomous Vehicles for Cyber-physical security. *IEEE International Conference on Systems*, Man, and Cybernetics, Banff, Canada, pp. 2655-2660.
DOI: 10.1109/SMC.2017.8123026

[39] Holbrow, C. H., Galvez, E. J., Parks, M. E. (2002). Photon quantum mechanics and beam splitters. *Am J Phys*, 70: 260–265.

[40] Galvez, E. J., Holbrow, C. H., Pysher, M. J., Martin, J. W., Courtemanche, N., Heilig, L., Spencer, J. (2005). Interference with correlated photons: five quantum mechanics experiments for undergraduates. *American Journal of Physics*, Volume 73, Issue 2, pp. 127-140.
DIO: 10.1119/1.1796811

[41] Galvez, E. J. (2005). Undergraduate Laboratories Using Correlated Photons: Experiments on the Fundamentals of Quantum Mechanics, *Innovative Laboratory Design*, pp. 113-118.

[42] Galvez, E. J., Beck, M. (2007). Quantum Optics Experiments with Single Photons for Undergraduate Laboratories. *Proc. SPIE 9665, Tenth International Topical Meeting on Education and Training in Optics and Photonics*, 966513, Ottawa, Ontario, Canada.
DIO: 10.1117/12.2207351.

[43] Dehlinger, D., Mitchell, M. W. (2002). Entangled photon apparatus for the undergraduate laboratory. *Am J Phys*, 70: 898–902.

[44] Thorn, J. J., Neel, M. S., Donato, V. W., Bergreen, G. S., Davies, R. E., Beck. M. (2004). Observing the quantum behavior of light in an undergraduate laboratory. *Am. J. Phys*, 72; 1210-1226.

[45] Conrad, A., Chaffee, D., Chapman, J., Chopp, C., Herdon, K., Hill, A., Sanchez-Rosales, D., Szabo, J., Gauthier, D. J., Kwia, P. G. (2019). Drone-based Quantum Key Distribution. *Bulletin*

of the American Physical Society, March Meeting, Volume 64, Number 2, Monday–Friday, March 4–8, Boston, Massachusetts, USA.

[46] Kent, A. (2012). Unconditionally Secure Bit Commitment by Transmitting Measurement Outcomes, *Physical Review Letters*, 109, 130501, 1-4.
DOI: 10.1103/PhysRevLett.109.130501

[47] Lunghi, T., Kaniewski, J., Bussieres, F., Houlmann, R., Tomamichel, M., Kent, A., Gisin, N., Wehner, S., Zbinden., H. (2013). Experimental bit commitment based on quantum communication and special relativity, *Physical Review Letters*, 111, 180504, pp. 1-5.

[48] Khoshnoud, F., Robinson, D., de Silva, C. W., Esat, I. I., Quadrelli, M. B. (2019). Research-informed service-learning in Mechatronics and Dynamic Systems. *The American Society for Engineering Education Conference*, Los Angeles, CA, April 4-6.

[49] Khoshnoud, F., Esat, I. I., de Silva, C. W., Quadrelli, M. B. (2019). Quantum Network of Cooperative Unmanned Autonomous Systems. *Unmanned Systems journal*, Vol. 07, No. 02, pp. 137-145.

[50] Khoshnoud, F., Esat, I. I., Quadrelli, M. B., Robinson, D., Edited by Clarence W. de Silva, C. W. (2020). Quantum Cooperative Robotics and Autonomy. *Instrumentation Journal*, in press.

[51] Khoshnoud, F., Esat, I. I., de Silva, C. W., Rhodes, J., Kiessling, A., Quadrelli, M. B. (2019). Self-powered Solar Aerial Vehicles: Towards infinite endurance UAVs. *Unmanned Systems*, Vol. 8, No. 2, 2020, pp. 1–23.

[52] Khoshnoud, F., McKerns, M., de Silva, C. W., Esat, I. I., Owhadi, H. (2017). Self-powered Dynamic Systems in the framework of Optimal Uncertainty Quantification, *ASME Journal of Dynamic Systems, Measurement, and Control*, Volume 139, Issue 9.

[53] Lin, S. H., Eyring, H. (1971). Solution of the Time-Dependent Schrodinger Equation by the Laplace Transform Method. *Proceedings of the National Academy of Sciences*, Vol. 68, No. 1, pp. 76-81.

[54] Sasakawa, T., J. (1966). Solution of Schrödinger Equation Involving Time. *Journal of Mathematical Physics*, 7, 721.

[55] Eyring, H., Kimball, G. E., Walters, J. (1944). Quantum Chemistry. John Wiley & Sons Inc; First edition, Hoboken, NJ.

[56] Galvez, E. J., Lecture notes, Chapter 19, Photons and Quantum Mechanics, Colgate University, NY.

[57] Kasap, S. O. (2013). Optoelectronics and Photonics: Principles & Practices. *Pearson Education*, Second Edition, Upper Saddle River, NJ, USA, ISBN-10: 0132151499.

[58] Townsend, J. S. (2000). A modern approach to quantum mechanics, University Science Books, Sausalito, California.

[59] https://www.thorlabs.com/ (accessed June 2019)

[60] https://www.parallax.com/ (accessed June 2019)

[61] https://store.dji.com/ (accessed June 2019)

[62] https://pixycam.com/ (accessed June 2019)

[63] Dehlinger, D., Mitchell. M. W. (2002). Entangled photons, nonlocality, and bell inequalities in the undergraduate laboratory, *Am J Phys*, 70: pp. 903–910.

[64] Bennett, C. H., Brassard, G., Crépeau, C., Jozsa, R., Peres, A., Wootters. William, K. (1993). Teleporting an Unknown Quantum State via Dual Classical and Einstein-Podolsky-Rosen Channels. *Physical Review Letters*, 70, 1895.

[65] Bouwmeester, D., Pan, J.-W., Mattle, K., Eibl, M., Weinfurter, H., Zeilinger, A. (1997). Experimental quantum teleportation. *Nature* 390: 575–579.

[66] Barrett, M. D., Chiaverini, J., Schaetz, T., Britton, J., Itano, W. M., Jost, J. D., Knill, E., Langer, C., Leibfried, D., Ozeri, R., Wineland, D. J. (2004). Deterministic quantum teleportation of atomic qubits, *Nature*, volume 429, pp. 737–739.

[67] Braunstein, S. L., Mann, A. (1995). Measurement of the Bell operator and quantum teleportation. *Physical Review A*, 51, R1727(R), 53, 630.

[68] Nölleke, C., Neuzner, A., Reiserer, A., Hahn, C., Rempe, G., Ritter, S. (2013). Efficient Teleportation Between Remote Single-Atom Quantum Memories. *Physical Review Letters,* 110, 140403.

[69] Bowmeester, D., Ekert, A., Zeilinger, A., (Eds.). (2000). The Physics of Quantum Information. *Springer*, Berlin.
ISBN 978-3-662-04209-0.

[70] Nielsen, M. A., Chuang, I. L. (2010). Quantum Computation and Quantum Information: 10th Anniversary Edition, *Cambridge University Press*, New York.
ISBN-13: 978-1107002173

[71] Yariv, A. (1989). Quantum Electronics, 3rd Edition, *John Wiley & Sons, inc*. USA. ISBN-13: 978-0471609971.

[72] An introduction video to Quantum Multibody Dynamics, Controls, Robotics, and Autonomy, https://www.youtube.com/watch?v=ForcnzWzG1M&t (accessed August 2019)

[73] Videos on demonstrations of the quantum robotic experiments: https://www.youtube.com/channel/UCai7HYeoCB-wgd9X3UlgDUA (accessed June 2019)

[74] Galvez, E. J. (2019). Colgate University, Email communications.

[75] Khoshnoud, F., Lamata, L., de Silva, C. W., Quadrelli, M. B. (2019). Quantum Teleportation for Control of Dynamical Systems and Autonomy, under review.

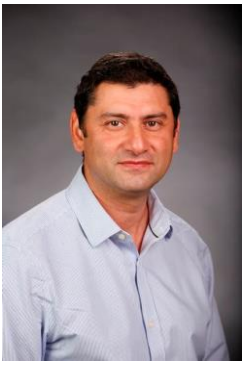

**Farbod Khoshnoud**, PhD, CEng, PGCE, HEA Fellow, is a faculty member in the college of engineering at California State Polytechnic University, Pomona, and a visiting associate in the Center for Autonomous Systems and Technologies, in the department of Aerospace Engineering at California Institute of Technology. His current research areas include Self-powered and Bio-inspired Dynamic Systems; Quantum Multibody Dynamics, Robotics, Controls and Autonomy, by experimental Quantum Entanglement, and Quantum Cryptography; and theoretical Quantum Control techniques. He was a research affiliate at NASA's Jet Propulsion Laboratory, Caltech in 2019; an Associate Professor of Mechanical Engineering at California State University, 2016-18; a visiting Associate Professor in the Department of Mechanical Engineering at the University of British Columbia (UBC) in 2017; a Lecturer in the Department of Mechanical Engineering at Brunel University London, 2014-16; a senior lecturer at the University of Hertfordshire, 2011-14; a visiting scientist and postdoctoral researcher in the Department of Mechanical Engineering at UBC, 2007-11; a visiting researcher at California Institute of Technology, 2009-11; a Postdoctoral Research Fellow in the Department of Civil Engineering at UBC, 2005-2007. He received his Ph.D. from Brunel University in 2005. He is an associate editor of the Journal of Mechatronic Systems and Control. He was the guest editor of the Quantum Engineering special issue of the Journal of Mechatronic Systems and Control.
Emails: fkhoshnoud@cpp.edu; farbodk@caltech.edu

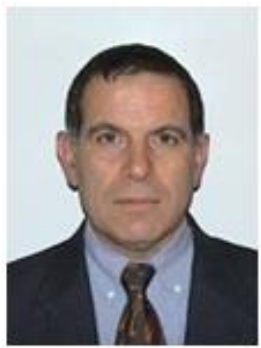

**Ibrahim I. Esat** has BSc and PhD from Queen Mary College of London University, after completing his PhD, he worked on mechanism synthesis at Newcastle university for several years, after this he moved to a newly formed University in Cyprus spending one and half year before returning back to UK joining to the University College of London University where he worked on designing "snake arm" robot. After this he moved to the Dunlop Technology division as a principal engineer working on developing CAD packages in particular surface modellers, following this he returned back to academia as a lecturer at Queen Mary College and later moving to Brunel University. He continued working closely with industry. He continued developing bespoke multi body dynamics software with user base in the UK, Europe and USA. Currently he is a full professor at Brunel University.
Email: Ibrahim.Esat@brunel.ac.uk

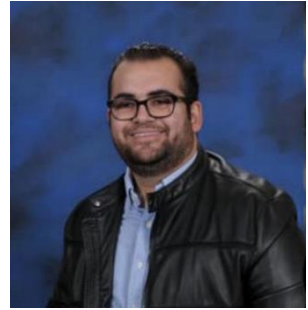

**Shayan Javaherian** received his B.S. in mechanical engineering from University of California, Berkeley. His passion is in the areas of soft robotics, controls, quantum robotics, and AI. He worked on underwater communication, ROVs, and laser tractions as his undergrad research at UC Berkeley and also as his graduate research at California State Polytechnic University, Pomona, which he is the project lead for the Intelligence & Autonomous Vehicles laboratory.
Email: sjavaherian@cpp.edu

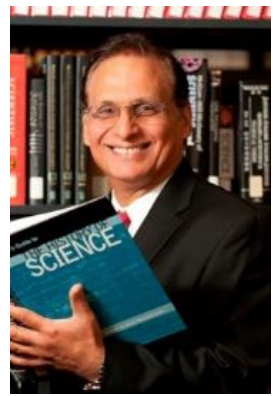

**Dr. Behnam Bahr** received Ph.D. in Mechanical Engineering from the University of Wisconsin-Madison in. His teaching and research are in the area of Biologically Inspired Robotics, Automation, and Autonomous Systems, Computer Aided Engineering, and Controls. He has authored and coauthored more than seventy journals and conference papers, and has been the advisor for nine Ph.D., and more than fifty master's degree students. He is currently the PI of the $2.6 million grant, Mentoring, Educating, Networking, and Thematic Opportunities for Research in Engineering and Science from the Department of Education. He served as the Associate Dean for Research and Graduate Studies from 2012-2016 in the College of Engineering at California State Polytechnique University-Pomona (Cal Poly Pomona). In that capacity he developed seminar series for faculty in the areas of Teaching, Research and Safety, and was instrumental in developing three new Master's emphasis in "System Engineering", "Environmental and Water Resource Engineering," and "Materials Engineering." He was also the Co-PI of the Cal Poly Pomona on the

Department of Education for the "First In The World" Program with the objective to flip the courses in the undergraduate STEM. Prior to joining the Cal Poly Pomona, he was the Boeing Endowed Manufacturing Professor at California State University Long Beach (CSULB), and in that capacity he developed the "National Center for Green Composite Technology", and was a key member for the $3 million NSF grant "Integrative Graduate Education and Research Traineeship Program between Arizona State University and CSULB. Dr. Bahr received FAA faculty summer fellowships from 1989 to 1992 at the FAA Technical Center in New Jersey working on various projects for the safety of the Aging Aircraft. He was the Chair of the Mechanical Engineering at Wichita State University 2005-2009, and was the PI/Co-PI of more than $3 million in external grants in the areas of manufacturing, robotics, and automation which were funded by the FAA and the aviation industry.
Email: bbahr@cpp.edu